\documentclass[12pt]{article}

\usepackage{amsmath, amssymb}
\usepackage{amsfonts}

\begin{document}


\def\a{\alpha}
\def\b{\beta}
\def\c{\varepsilon}
\def\d{\delta}
\def\e{\epsilon}
\def\f{\phi}
\def\g{\gamma}
\def\h{\theta}
\def\k{\kappa}
\def\l{\lambda}
\def\m{\mu}
\def\n{\nu}
\def\p{\psi}
\def\q{\partial}
\def\r{\rho}
\def\s{\sigma}
\def\t{\tau}
\def\u{\upsilon}
\def\v{\varphi}
\def\w{\omega}
\def\x{\xi}
\def\y{\eta}
\def\z{\zeta}
\def\D{{\mit \Delta}}
\def\G{\Gamma}
\def\H{\Theta}
\def\L{\Lambda}
\def\F{\Phi}
\def\P{\Psi}
\def\S{\Sigma}
\def\V{\varPsi}
\newcommand{\EV}{ \,{\rm eV} }
\newcommand{\KEV}{ \,{\rm keV} }
\newcommand{\MEV}{ \,{\rm MeV} }
\newcommand{\GEV}{ \,{\rm GeV} }
\newcommand{\TEV}{ \,{\rm TeV} }

\def\o{\over}
\newcommand{\sla}[1]{#1 \llap{\, /}}
\newcommand{\beq}{\begin{eqnarray}}
\newcommand{\eeq}{\end{eqnarray}}
\newcommand{\gsim}{ \mathop{}_{\textstyle \sim}^{\textstyle >} }
\newcommand{\lsim}{ \mathop{}_{\textstyle \sim}^{\textstyle <} }
\newcommand{\vev}[1]{ \left\langle {#1} \right\rangle }
\newcommand{\bra}[1]{ \langle {#1} | }
\newcommand{\ket}[1]{ | {#1} \rangle }
\newcommand{\1}{\mbox{1}\hspace{-0.25em}\mbox{l}}

\newcommand{\cM}{{\cal M}}
\newcommand{\cB}{{\cal B}}


\baselineskip 0.7cm

\begin{titlepage}

\begin{flushright}
YITP-08-97 \\
\end{flushright}

\vskip 1.35cm
\begin{center}
{\large \bf
Strongly Coupled Semi-Direct Mediation \\
of Supersymmetry Breaking
}
\vskip 1.2cm
M.~Ibe$^1$,~Izawa K.-I.$^{2,3}$,~and Y.~Nakai$^2$
\vskip 0.4cm

{\it $^1$SLAC National Accelerator Laboratory, Menlo Park,\\ CA 94025, USA}

{\it $^2$Yukawa Institute for Theoretical Physics, Kyoto University,\\
     Kyoto 606-8502, Japan}

{\it $^3$Institute for the Physics and Mathematics of the Universe, University of Tokyo,\\
     Chiba 277-8568, Japan}

\vskip 1.5cm

\abstract{
Strongly coupled semi-direct gauge mediation models
of supersymmetry breaking through massive mediators
with standard model charges
are investigated by means of composite degrees of freedom.
Sizable mediation is realized to generate the standard model
gaugino masses for a small mediator mass without breaking
the standard model symmetries.
}
\end{center}
\end{titlepage}

\setcounter{page}{2}

\section{Introduction}

Supersymmetry (SUSY)
\cite{luty} is expected to be a crucial ingredient
of basic laws in Nature.
It is an attractive possibility that SUSY is  broken
at low energy within the experimental reach in the near future.
Among others, low-energy dynamics with gauge mediation
between a hidden sector of SUSY breaking and the visible sector
of SUSY standard model may be phenomenologically viable
\cite{Giu}.
In particular, the gauge interactions are flavor blind,
so that the unwanted flavor-changing processes are naturally suppressed.\footnote{The SUSY flavor problem is also discussed in \cite{luty}.}

The pivot of gauge mediation consists of messenger
fields that are charged under the standard model gauge symmetries.
We can assume messengers of three types
\cite{Sei}:
minimal, direct, and semi-direct
\cite{Iza,Yan}.%
\footnote{Some other ingredients than messenger fields may well be present
such as additional nonabelian messenger gauge interaction
\cite{Iza,Ran}, which we do not consider in this paper.}
Let $X=m+\theta^2 F$ be a representative spurion
for R and SUSY breaking (presumably with dynamical origin),
where $\theta$ denotes the superspace coordinate.
The minimal gauge mediation is given by a superpotential term
$Xq{\tilde q}$ with a standard model vector-like pair $q$ and ${\tilde q}$
of chiral superfields as the messengers.
The direct gauge mediation is more economical class which is given by a superpotential term
$XQ{\tilde Q}$ with standard model vector-like pairs $Q$ and ${\tilde Q}$
of chiral superfields as the mediators with hidden gauge interaction charges
(whose dynamics cause SUSY breaking encoded in the $X$ value).
The semi-direct mediation which, as we shall see shortly, can overcome one of the difficulties of the direct gauge mediation is given by a superpotential mass term
$\mu Q{\tilde Q}$ with $\mu$ a constant and a representative
term $X\F {\tilde \F}$
with a hidden gauge interaction vector-like pair
$\F$ and ${\tilde \F}$ of standard-model singlet chiral superfields.%
\footnote{This by no means gives a complete classification
of gauge mediation models. For example, the models in
Ref.\cite{Nom}
contain massive messengers without hidden gauge interaction charges.} 

An important difference of the semi-direct gauge mediation from
the direct mediation
is that the messengers do not play important roles
in dynamical SUSY breaking.
Thus, in the semi-direct mediation models, the rank of gauge group
required in the SUSY breaking sector can be smaller than
that in the direct mediation models. 
Hence, in the semi-direct mediation model,
we can ameliorate the Landau pole problem that is often encountered
in the direct mediation models at low-energy scale.

In the semi-direct mediation, however, it seems problematic that 
the gaugino masses in the SUSY standard model vanish to the leading order
in $F$ (the gaugino screening\,\cite{Ark}). This is because the supersymmetric masses of the messengers do not lead to the SUSY breaking in the holomorphic standard-model gauge coupling and the leading contribution only occurs by the wavefunction
renormalization at the higher loop order.
On the other hand, the scalar masses emerge at the leading order in $F$.
Thus, the gaugino masses are suppressed at least by $|F/m^2|^2$ compared with
the scalar masses,%
\footnote{The $|F/m^2| \lsim 1$ is required for the messenger sectors not to have
tachyonic modes.}
which leads to a little hierarchy between the gaugino masses and the scalar masses
without careful tuning between the sizes of $F$ and $m$. If this hierarchy exists, we cannot set the gaugino masses and the scalar masses at the order $1$ TeV at the same time, which causes the standard model hierarchy problem again.

In this paper, we consider a possible amelioration of the little hierarchy 
between the gaugino masses and the scalar masses in the 
semi-direct mediation while keeping the indirectness of the messengers--SUSY breaking 
interaction. 
For that purpose, we introduce a new mechanism of mediating the SUSY breaking effect to the SUSY standard model, that is, we consider a model of {\sl strongly coupled} semi-direct
gauge mediation where the mass of the mediators are small compared with the dynamical scale 
in the SUSY breaking sector.
In the strongly coupled model, the semi-direct messengers make bound states 
due to the strong dynamics in the SUSY breaking sector and obtain non-vanishing SUSY breaking 
vacuum expectation values, which make it possible for the gaugino masses 
in the SUSY standard model to have the leading contribution in $F$.

In the next section, we provide our model of strongly coupled
semi-direct gauge mediation and derive its effective theory description.
In section 3, the SUSY-breaking vacuum of the model is investigated in terms of
the effective theory. Section 4 is devoted to estimating the standard
model gaugino masses to see the scale of visible soft masses.
The final section concludes the paper with a few comments.

\section{The model}

We adopt hidden SUSY $SU(N)$ gauge theory with a superpotential
\beq
 W=X(\F {\tilde \F})+\mu {\, \rm tr} Q{\tilde Q}
 + \kappa \, ({\rm tr} Q{\tilde Q})^2,
 \label{model}
\eeq
where the spurion $X$ is given by $X=m + \theta^2 F$,
$\F$ and ${\tilde \F}$
denote a vector-like pair of the hidden gauge interaction,
and $Q$ and ${\tilde Q}$ are (anti-)fundamentals with $N_f=N+1$
massive flavors.
The subgroups of the flavor symmetry, $SU(N_f)$, are eventually identified
with the gauge groups in the standard model.
In particular, $N=4$ implies $N_f=5$ with $SU(N_f)$
grand unification structure of the visible sector.
In the above superpotential, $\mu$ denotes a mass parameter of $Q$'s which
is assumed to be smaller than the dynamical scale of $SU(N)$ gauge theory, $\L_H$, {\it i.e.}
$\m \ll \L_H$, and $\k$ denotes a coupling constant with a negative mass dimension.

Let us first integrate out the hidden matter $\F$, ${\tilde \F}$, assuming $m\gg \L_H$,
which results in SUSY QCD-like theory of $N_f$ massive flavors
with mass $\mu$ and spurious dynamical scale $\L^b=X^a\L_H^{b-a}$,
where $b=3N-N_f=2N-1$ and $X$ is again $X = m+\h^2F$.
(For example, $a=N_\F$ for $N_\F$ pairs of (anti-)fundamentals
$\F$ and ${\tilde \F}$.)
For later convenience, we define
\beq
\L=\L_0(1+\theta^2 R)=m^{a \o b}\L_H^{1-{a \o b}}
   \left( 1+\theta^2 {aF \o bm} \right).
 \label{scale}
\eeq
This ratio $R \sim F/m \gtrsim 100 \, \mathrm{TeV}$ turns out to be a characteristic mass scale
of gauge mediation.

Further, we proceed to integrating out the gauge and matter degrees of
freedom to obtain the effective theory of meson $M$ and baryon
$B$, ${\tilde B}$ chiral superfields.
Under the standard model gauge groups, the mesons behave as 
the singlet and adjoint representations, 
while the baryons behave as the (anti-)fundamental representations.%
\footnote{The meson $M$ is an $N_f \times N_f$ matrix
and the baryons $B$, ${\tilde B}$ are $N_f$-component vectors.}
The superpotential and K\"ahler potential of those composite fields are given by\,\cite{Kap}
\beq
 W_{eff}={1 \o \L^{b}}(BM{\tilde B}-\det M)+\mu {\, \rm tr} M
 + \kappa \, ({\rm tr} M)^2,
\eeq  
and 
\beq
 K_{eff}=\left( {\a^2 |M|^2 \o |\L|^2} + {\b^2 |B|^2 \o |\L|^{2(N-1)}}
         + {\b^2|{\tilde B}|^2 \o |\L|^{2(N-1)}} \right) {K},
\eeq
where $\a$ and $\b$ are positive constants and
\beq
 {K}({M/\L^2}, {B/\L^{N}}, {{\tilde B}/\L^{N}}, {\L/X}, {\mu/\L}, \k \L)
 = 1 + \cdots.
\eeq
Here, the ellipsis denotes the higher contributions to the
K{\" a}hler potential in the arguments.

In the supersymmetric limit, $F= 0$, with $\mu, \k >0$,
the vacuum expectation values are given by
$\vev{M} \simeq \sqrt[N]{\mu \L_0^b} \1$ and $\vev{B}=\tilde{\vev{B}}={\mathbf 0}$,
where we have assumed the coupling constant $\kappa$ to be so small
as $\k \! \sqrt[N]{\mu \L_0^b} \ll \mu$.
We presume the case $|F| \ll m^2$ and $\mu \ll \L_0$
to adopt an approximation $K \simeq 1$ in the following sections.

\section{The SUSY-breaking vacuum}

Now, let us investigate how the original SUSY breaking effect in $X$
is propagated into the mediator fields.
For small SUSY breaking $|R| \ll \mu$,
we still expect $\vev{B}=\tilde{\vev{B}}={\mathbf 0}$
and diagonal $\vev{M}$, and hence,
the standard model gauge symmetries are not broken at the vacuum.
The (soft) masses of the composite degrees of freedom may be seen
from Eqs.(\ref{superp}) and (\ref{kahler})
for retrospective justification of these vacuum expectation values.

To see the SUSY breaking effects on composite fields explicitly, 
let us define the normalized chiral superfields:
\beq
 {\cal M}=\a M/\L_0, \quad {\cal B}=\b B/\L_0^{N-1},
  \quad {\tilde {\cal B}}=\b {\tilde B}/\L_0^{N-1}.
\eeq
With the aid of
\beq
 {1 \o \L^b}={1 \o \L_0^{b}}\left( 1-\theta^2 {bR} \right)
 ={1 \o \L_0^{b}}\left( 1-\theta^2 {aF \o m} \right),
\eeq
we obtain
\beq
 W_{eff} = \left( 1-\theta^2 {bR} \right)
 \left( \l \cB {\cM}{\tilde \cB} - {\g \o \L_0^{N-2}} \det {\cM} \right)
 + {\tilde \mu} \L_0 \,{\rm tr} \cM + {\tilde \kappa} \L_0^2 \, ({\rm tr} \cM)^2,
\eeq
where $\l=\a^{-1} \b^{-2}$, $\g=\a^{-(N+1)}$, ${\tilde \mu}=\a^{-1}\mu$, ${\tilde \kappa}=\a^{-2}\kappa$
and
\beq
 K_{eff} \simeq \left| 1-\theta^2 R \right|^2 |\cM|^2
         + \left| 1-\theta^2 {(N-1)R} \right|^2
         (|\cB|^2 + |{\tilde \cB}|^2).
\eeq

The effective scalar potential is obtained by performing the integrals over the
Grassmann coordinates in the above effective superpotential and the K{\"
a}hler potential. Then, minimizing this effective potential, we see
\beq
\label{F}
 \frac{\cM_1}{\cM_0} \simeq {bR \o N}
 \left(1 + \frac{2\kappa \vev{{\rm tr} M}}{N\mu} \right)\propto F,
\eeq
for the vacuum expectation value
$ \vev{\cM} = (\cM_0+\theta^2\cM_1) \1$
to the leading order of the SUSY-breaking parameter $R$ of gauge mediation,
where we abuse the notation $\vev{{\rm tr} M}$ for its lowest component
as $\vev{{\rm tr} M} \simeq N_f \! \sqrt[N]{\mu \L_0^b}$.

Therefore, we found that the diagonal component of the meson fields, ${\cal M}\propto \1$, 
obtains a non-vanishing SUSY breaking $F$-term vacuum expectation value.
As we will see below, the $F$-term vacuum expectation value of the diagonal component 
plays a crucial role to generate the gaugino mass in the SUSY standard model 
at the leading order in $F$.

\section{Visible gaugino masses}

Now, we  evaluate the standard model gaugino masses in terms of the vacuum expectation value discussed above.
In the SUSY-breaking vacuum,
we obtain
\beq
 W_{eff} \simeq \left( 1-\theta^2 {bR} \right)
         \left( \l \cB \vev{\cM}{\tilde \cB}
  + {1 \o 2} {\g \o \L_0^{N-2}}
  (\cM_0+\theta^2\cM_1)^{N-1}\,{\rm tr}{\tilde \cM}^2 \right),
 \label{superp}
\eeq
where ${\tilde \cM}= \cM-\vev{\cM}$, and
\beq
 K_{eff} \simeq \left| 1-\theta^2 {R} \right|^2 |{\tilde \cM}|^2
         + \left| 1-\theta^2 {(N-1)R} \right|^2
         (|\cB|^2 + |{\tilde \cB}|^2),
 \label{kahler}
\eeq
as the leading approximations to the quadratic terms
of the composite fields charged under the standard model gauge symmetries.

There are two main contributions to the gaugino masses.\footnote{The SUSY breaking effects arising from the K\"ahler potential, while present, are subleading in $F$ and therefore will not be considered in this paper.}
One is from the threshold corrections which emerge when we integrate out
the mesons $\cal M$ and the baryons $\cal B $ and $\tilde {\cal B}$  which obtain 
SUSY breaking scalar masses from the $F$-term vacuum expectation 
value of the diagonal $\cal M$ via the superpotential in Eq.\,(\ref{superp}).
The other contribution at the leading order comes from the threshold corrections
which emerge when we integrate out the heavier modes than the mesons and the baryons 
in the strong dynamics.
Although, it is highly difficult to calculate the threshold corrections from those 
heavy modes in the strong dynamics in general, it is possible to extract the gaugino mass
with the help of the anomaly matching conditions of the global symmetries\,\cite{Nakai}.

First, let us calculate the gaugino mass contribution from the mesons and baryons by analytic continuation of the gauge coupling renormalization factors into superspace\,\cite{Rat}.
By using the superpotential Eq.\,(\ref{superp}) and the vacuum expectation value 
of $F$-term in Eq.\,(\ref{F}), we obtain the gaugino mass contribution from mesons and baryons,
\beq
 \d m_{1/2} \simeq {\a_{\rm sm} \o 2\pi} bR
 \left(\frac{\kappa \vev{{\rm tr} M}}{\mu} -1 \right)\propto F,
\eeq
where $\a_{\rm sm}$ indicates the standard model gauge couplings
(squared over $4\pi$) and $b$ and $R$ are defined in Eq.(\ref{scale}) and
just above it.

The other leading contribution to the gaugino masses, {\it i.e.}, 
the threshold effects of the heavier modes in the strong dynamics, 
is estimated in the following way.
Let us first recapitulate the gauge and global symmetries of the model,
whose charge assignments to the matter fields in the microscopic and the
macroscopic theory are shown in tables 1 and 2, respectively.
We also assign charges to the coupling constants, regarding them as background chiral superfields.
The anomalous $U(1)_A$ global symmetry can be treated as a symmetry by rotating the dynamical scale $\Lambda$.

\begin{table}[tdp]
\begin{center}
\begin{tabular}{c|ccccc}
 & $SU(N)$ & $SU(N_f)$
 & $U(1)_R$ & $U(1)_A$
 \\
 \hline
$Q$ & $\mathbf N$ & $\mathbf {N}_f$
& $1-\frac{N}{N_f}$ & $1$
\\
  $\tilde{Q}$  &  $\mathbf  {\overline{N}}$ & $\mathbf  {\overline{N}}_f$
  & $1-\frac{N}{N_f}$ & $1$
  \\
  $\mu$ & {\bf 1} & {\bf 1} & $2\frac{N}{N_f}$ & $-2$ & \\
  $\kappa$ & {\bf 1} & {\bf 1} & $-2 + 4\frac{N}{N_f}$ & $-4$ & \\
  $\Lambda^{b}$ & {\bf 1} & {\bf 1} & $0$ & $2N_f$
\end{tabular}
\caption{the charge assignments in the microscopic theory.}
\end{center}
\label{tab:micro}
\end{table}

\begin{table}[tdp]
\begin{center}
\begin{tabular}{c|ccccc}
 & $SU(N_f)$
 & $U(1)_R$ & $U(1)_A$
 \\
 \hline
$B$ & $\mathbf {N}_f$
& $N-\frac{N^2}{N_f}$ & $N$
\\
  $\tilde{B}$  & $\mathbf  {\overline{N}}_f$
  & $N-\frac{N^2}{N_f}$ & $N$
  \\
  $M$ & {\bf adj + 1} & $2-2\frac{N}{N_f}$ & $2$ & \\
\end{tabular}
\caption{the charge assignments in the macroscopic theory.}
\end{center}
\label{tab:macro}
\end{table}

The $U(1)_A$--$SU(N_f)^2$ anomaly matching condition implies that we further need an effective operator of the form
$(\ln \, \L^b) \, {\cal W}^\a {\cal W}_\a$,
where $\cal W$ denotes the field strength superfield of the standard
model gauge multiplets.
Hence we conclude that the standard model gaugino masses are given by
\beq
 m_{1/2} \simeq {\a_{\rm sm} \o 2\pi} bR
 \left(\frac{\kappa \vev{{\rm tr} M}}{\mu} -1 \right) + {\a_{\rm sm} \o 2\pi} bR
 = {\a_{\rm sm} \o 2\pi} {a\kappa F \o \mu m}\vev{{\rm tr} M} \propto F,
 \label{gaugino}
\eeq
where the $\k$-independent contributions cancel out and the results are proportional to the dynamical scale $\L_0$ of the theory in accord with the perturbative gaugino screening.
As a result, we find that the gaugino mass in the standard model can be 
generated at the leading order in $F$ which comes from the non-vanishing 
$F$-term expectation value of the diagonal component of the composite meson.

\section{Conclusion}

In this paper, we proposed a possible solution of the gaugino screening problem in the 
semi-direct mediation by introducing the strongly coupled messenger sector. We investigated the model Eq.(\ref{model}) of strongly coupled semi-direct
gauge mediation and its effective theory
given by Eqs.(\ref{superp}) and (\ref{kahler}).
Sizable mediation is realized to generate the standard model
gaugino masses Eq.(\ref{gaugino}) without breaking
the standard model symmetries. That is, the leading contribution to the gaugino masses does not vanish and is proportional to the dynamical scale of the model with a small mediator mass.
We note that similar analyses can be performed for more general gauge theories,
though we have definitely considered the model of hidden $SU(N)$ gauge theory
with $N_f=N+1$.

Finally let us mention a few obvious issues to be investigated.
One is the scale of the gravitino mass. It depends on the origin of
SUSY-breaking spurion,
which we have not specified in this paper.
Another is the SUSY-breaking scale itself. We have restricted ourselves
to the case of small SUSY breaking, $|F| \ll m^2$, though 
large SUSY-breaking case seems interesting to explore,
which may achieve lowest-scale gauge mediation with 
a cosmologically favorable tiny gravitino mass.

\section*{Acknowledgements}

This work was supported by the Grant-in-Aid for Yukawa International
Program for Quark-Hadron Sciences, the Grant-in-Aid
for the Global COE Program "The Next Generation of Physics,
Spun from Universality and Emergence", and
World Premier International Research Center Initiative
(WPI Initiative), MEXT, Japan.



\end{document}